# Stable Orbits of Rigid, Rotating, Precessing, Massive Rings

Edward D. Rippert
*Whitehawk Systems, 3 Whitehawk Circle, Boise, ID, 83716, USA.*
rippert@whitehawksystems.com

**The dynamics of a rigid, rotating, precessing, massive ring orbiting a point mass within the perimeter of the ring are considered. It is demonstrated that orbits dynamically stable against perturbations in three dimensions exist for a range of rigid body rotation parameters of the ring. Previous analysis and some well-known works of fiction have considered the stability of both rigid and flexible, non-precessing ring systems and found that they are unstable in the plane of the ring unless an active stabilization system is employed. There does not appear to be any analyses previously published considering rigid body precession of such a system or that demonstrate passive stability in three dimensions. Deviations from perfect rigidity and possible applications of such a system are discussed.**

Keywords: Orbital Ring, Ring World, Stability, Precession, Magic Angle

## I. Introduction

Solid rings rotating about a central mass have been considered for various applications ranging from a ring of geosynchronous satellites around the Earth [1] (Polyakov-Ring) to fictional accounts of a gigantic artificial habitat around a star rotating to provide artificial gravity [2] (Niven-Ring). A Polyakov-Ring rotates at or slightly above the orbital speed at its distance from the central mass. As such it lacks significant tension on the ring and is a flexible structure. It has been shown that such a structure is unstable against perturbations in the plane of the ring but can be stabilized by a feedback control system controlling the local length of ring segments [3]. A Niven-Ring is provided with rigidity due to the high rate of spin needed to produce artificial gravity. It has been shown that a rigid Niven-Ring is also unstable against perturbations in the plane of the ring but is stable against perturbations perpendicular to the ring [4]. It has been suggested that a Niven-Ring can also be stabilized by a feedback control system consisting of radial thrusters on the ring [5]. Another concept for an orbital ring system consisting of a Low Earth Orbit, massive ring electromagnetically tethered to geostationary skyhooks has also been published [6 – 8] but represents a different class of object, which requires a large, continuous power consumption to remain stable.

The analyses in [1] and [3] assume that the ring mass, $M_R$, is much smaller than the central point mass, $M_S$ ($M_R << M_S$). Another analysis assumes that $M_R >> M_S$ and discusses stable and unstable orbits of a point mass about a stationary, massive ring [9].

This paper addresses the orbital stability of a thin, rigid, circular ring with a point mass within the perimeter of the ring. It is assumed that the central point mass is much greater than the mass of the ring, $M_R << M_S$. A combination of examination of the gravitational potential and numerical simulations is used to determine possible stable orbits of this system as a function of rigid body rotation parameters of the ring. It is shown that passively stable orbits do exist for a range of rigid body rotation parameters and that these orbits are stable in three dimensions. An introductory, qualitative discussion of the effects of imperfect rigidity and imperfect elasticity and a discussion of some speculative applications are provided.

## II. Gravitational Potential Energy

In order to examine the stability of orbits and derive the equations of motion, the gravitational potential energy of the ring and point mass system is derived following the methodology of [9]. Consider a uniform density, $\rho$, thin ring of mass $M_R$ and a point mass of mass $M_S$. Now define a body fixed, Cartesian coordinate system with basis vectors $e_1$, $e_2$ and $e_3$, origin at the center of the ring, C. Define the vector from C to the point mass, $X$ and the radial vector from C to a differential mass element, $dm$, on the ring is $R$ with magnitude $R$. Now define the following scalars: In cylindrical coordinates with origin at C, the radius of $X$ in the plane of the ring is d and the height of $X$ perpendicular to the ring is $h$. The distance from the point mass to the differential mass element, $dm$, is $r$. The angle between $d$ and $R$ is $\theta$. The shortest distance from the point mass to the ring is $q$. The longest distance from the point mass to the ring is $p$. Figure 1 provides a schematic diagram of the massive ring and point mass system.

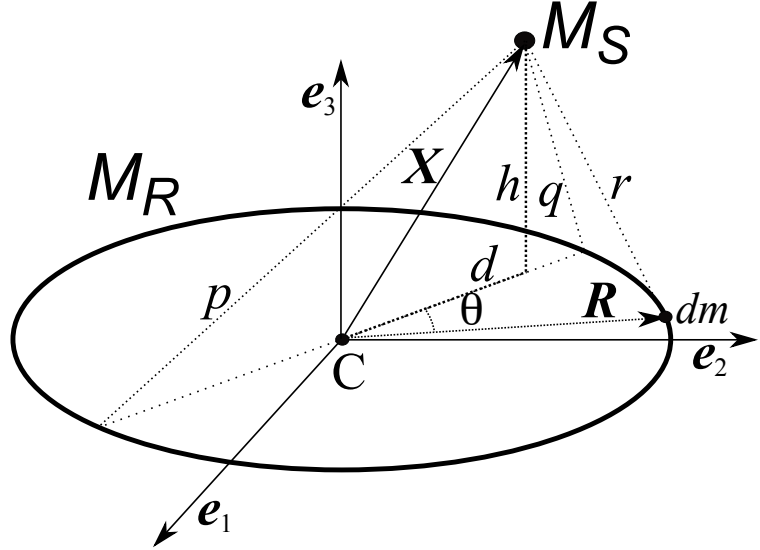

**Figure 1: Schematic Diagram of a Massive Ring – Point Mass System**

Now since $dm = \rho R d\theta$ and $M_R = 2\pi\rho R$, then the potential Energy, $U$, is

$$U = -GM_S \int_{-\pi}^{\pi} \frac{dm}{r} = -GM_S\rho R \int_{-\pi}^{\pi} \frac{d\theta}{r} = -\frac{GM_R M_S}{\pi} \int_0^{\pi} \frac{d\theta}{r} \tag{1}$$

Note that (1) takes advantage of the symmetry in the problem to change the integration range from $-\pi$ to $\pi$, to 0 to $\pi$ and added an extra factor of 2 outside the integral. Now note that

$$p^2 = (d+R)^2 + h^2; \quad q^2 = (d-R)^2 + h^2 \tag{2}$$

such that

$$p^2 + q^2 = 2\left(d^2 + R^2 + h^2\right); \quad p^2 - q^2 = 4Rd \tag{3}$$

and

$$r^2 = (d - R\cos\theta)^2 - R^2\sin^2\theta + h^2 = d^2 + R^2 + h^2 - 2Rd\cos\theta. \tag{4}$$

Combining (2), (3) and (4) and utilizing a half angle formula one can write a new expression for *r*,

$$r^2 = p^2\left[1 - \left(\frac{p^2 - q^2}{p^2}\right)\cos^2\frac{\theta}{2}\right]. \tag{5}$$

Now defining

$$k^2 = \frac{p^2 - q^2}{p^2} < 1 \qquad (6)$$

and

$$\varphi = \frac{\pi - \theta}{2}, \qquad (7)$$

the potential energy (1) can be written as

$$U = -\frac{2GM_S M_R}{p\pi} \int_0^{\frac{\pi}{2}} \frac{d\varphi}{\sqrt{1 - k^2 \sin^2 \varphi}} = -\frac{2GM_S M_R}{p\pi} K\left(k^2\right), \qquad (8)$$

where K is the Complete Elliptic Integral of the first kind. Finally $U$ can be written in terms of $d$ and $h$

$$U(d, h) = -\frac{2GM_S M_R}{\pi\sqrt{R^2 + 2dR + d^2 + h^2}} K\left(\frac{4Rd}{R^2 + 2dR + d^2 + h^2}\right). \qquad (9)$$

In order to examine the form of the potential energy in (9), the dimensionless variables $\xi = d/R$ and $\eta = h/R$ and the normalized potential $u = U/(G M_S M_R/R)$ are defined such that

$$u(\xi, \eta) = -\frac{2}{\pi\sqrt{1 + 2\xi + \xi^2 + \eta^2}} K\left(\frac{4\xi}{1 + 2\xi + \xi^2 + \eta^2}\right). \qquad (10)$$

Note that $\xi \geq 0$ since $d$ is a radius in cylindrical coordinates.

Figure 2 is a surface/contour plot of $u$ in the $\xi$–$\eta$ plane. The stability along the $\xi = 0$ axis and the instability along the $\eta = 0$ axis, as pointed out in [4], is clear from the shape of the potential. Note that the definition of $\xi$ requires that it be nonnegative, and Fig. 2 shows a reflection of the data about the $x$ axis to make interpretation of the potential more intuitive. This convention shall be used widely in this paper.

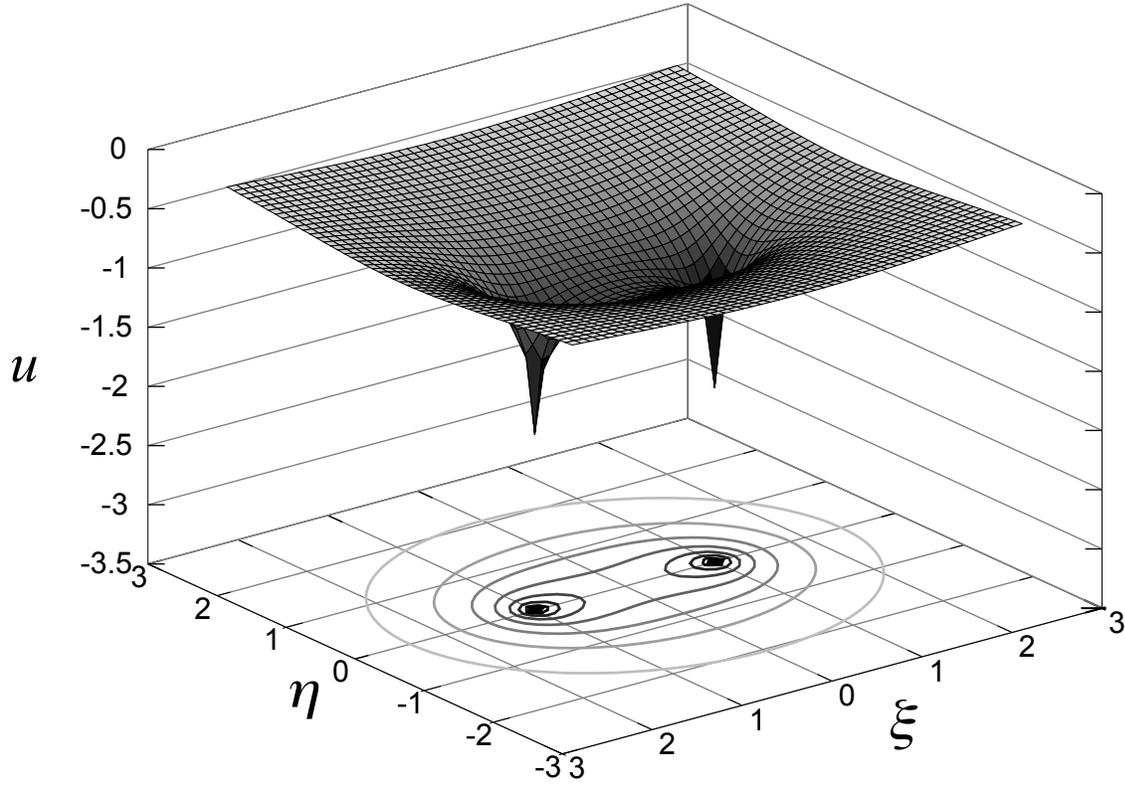

**Fig. 2: Surface/Contour Plot of the Normalized Gravitational Potential, *u*, in the $\xi$–$\eta$ Plane**

Now consider the ring and point mass system when the rigid ring has an angular momentum ***L***. If ***L*** is parallel to the ***e***$_3$ axis, then the conclusions of [4] remain valid. If ***L*** is not parallel to the ***e***$_3$ axis, then the ring will be precessing and there will be a varying gravitational potential. Also, gravitational force that does not act on the symmetry axes of the ring can be expected to have a torque, which will introduce gyroscopic effects into the orbital dynamics.

Consider a rotating, precessing, ring that has a precessional period that is small compared to the time scale of its orbital motion. The effective gravitational potential of this rapidly precessing ring can be approximated by averaging the normalized gravitational potential, (10), over one precessional period.

In order to average over the precessional period of the ring, consider the torque free solution to a rotating symmetric rigid body [10]. The total angular momentum, ***L***, will be a constant in a fixed spatial reference frame and the ***e***$_3$ axis of the ring will trace out a precessional cone about ***L*** at a fixed half-angle, $\alpha$, at a constant rate. The gravitational potential's time average of one period of this motion will therefore be the average for all orientations of the ring as it traces the cone. Define the unit vector parallel to $L$ as $e_L$ and the precessional angle about $e_L$ as $\beta$, as in Fig. 3.

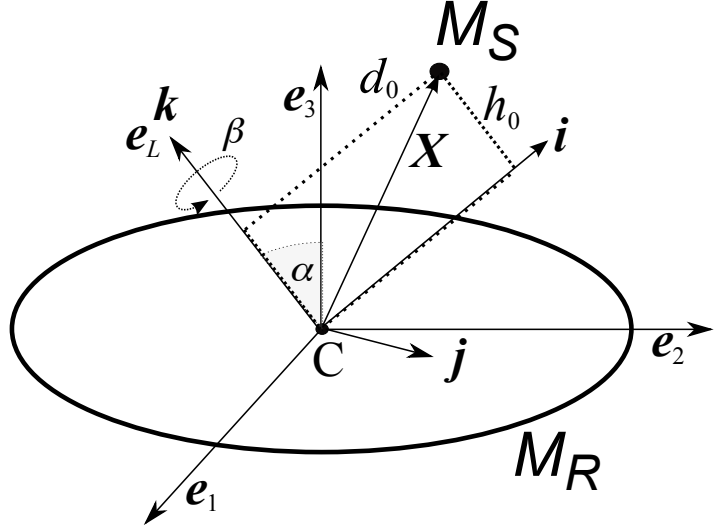

Figure 3: Definitions of precession angle, $\beta$ and precessional cone half-angle, $\alpha$, relative to the ring and point mass system. Note that the 2-D projection used contains a binary state optical illusion on the relative configuration of the *1-2-3* and *i-j-k* axes, this is resolved by the definition that *i-j-k* is the first permutation of an orthogonal, right handed coordinate system, as is *1-2-3*.

Now consider a coordinate system with origin at C and basis vectors $i, j$ and $k$, as in Fig. 3 with $k = e_L$, such that $X = id_0 + kh_0$. Then define the functions $d'(\alpha, \beta, d_0, h_0)$ and $h'(\alpha, \beta, d_0, h_0)$ as the in the-plane-of-the-ring and perpendicular to-the-plane-of-the-ring components of $X$, respectively, for a given $\alpha$ and $\beta$ (the same scalar values as $h$ and $d$ from Fig. 1). Then, by analogy with spherical coordinates with $e_3$ as the radial unit vector, $\beta$ as the azimuthal angle and $\alpha$ as the polar angle, it is possible to write the orthogonal, right handed coordinate system

$$\begin{aligned} e_\alpha(\alpha, \beta) &= i\cos\beta\cos\alpha + j\sin\beta\cos\alpha - k\sin\alpha \\ e_\beta(\alpha, \beta) &= -i\sin\beta + j\cos\beta \\ e_3(\alpha, \beta) &= i\cos\beta\sin\alpha + j\sin\beta\sin\alpha + k\cos\alpha \end{aligned} \quad (11)$$

Since $h' = X \cdot e_3$ and $\|X\|^2 = d'^2 + h'^2$,

$$\begin{aligned} h'(\alpha, \beta, d_0, h_0) &= d_0\cos\beta\sin\alpha + h_0\cos\alpha \\ d'(\alpha, \beta, d_0, h_0) &= \sqrt{d_0^2 + h_0^2 - (d_0\cos\beta\sin\alpha + h_0\cos\alpha)^2} \end{aligned} \quad (12)$$

By defining the dimensionless variables $\xi_0 = d_0 / R$ and $\eta_0 = h_0 / R$, one can write the dimensionless form of (12)

$$\eta'(\alpha, \beta, \xi_0, \eta_0) = \xi_0 \cos\beta \sin\alpha + \eta_0 \cos\alpha$$
$$\xi'(\alpha, \beta, \xi_0, \eta_0) = \sqrt{\xi_0^2 + \eta_0^2 - (\xi_0 \cos\beta \sin\alpha + \eta_0 \cos\alpha)^2}. \tag{13}$$

Using (13) and (10) one can write an expression for an averaged potential over the precessional period.

$$u_{AV}(\alpha, \xi_0, \eta_0) = \frac{1}{2\pi} \int_0^{2\pi} u\left(\xi'(\alpha, \beta, \xi_0, \eta_0), \eta'(\alpha, \beta, \xi_0, \eta_0)\right) d\beta. \tag{14}$$

Figure 4 is a plot of the numerically calculated curvature of the averaged potential, $u_{AV}$, at the coordinates $\xi_0 = 0$ and $\eta_0 = 0$, as a function of the precessional cone half-angle, $\alpha$.

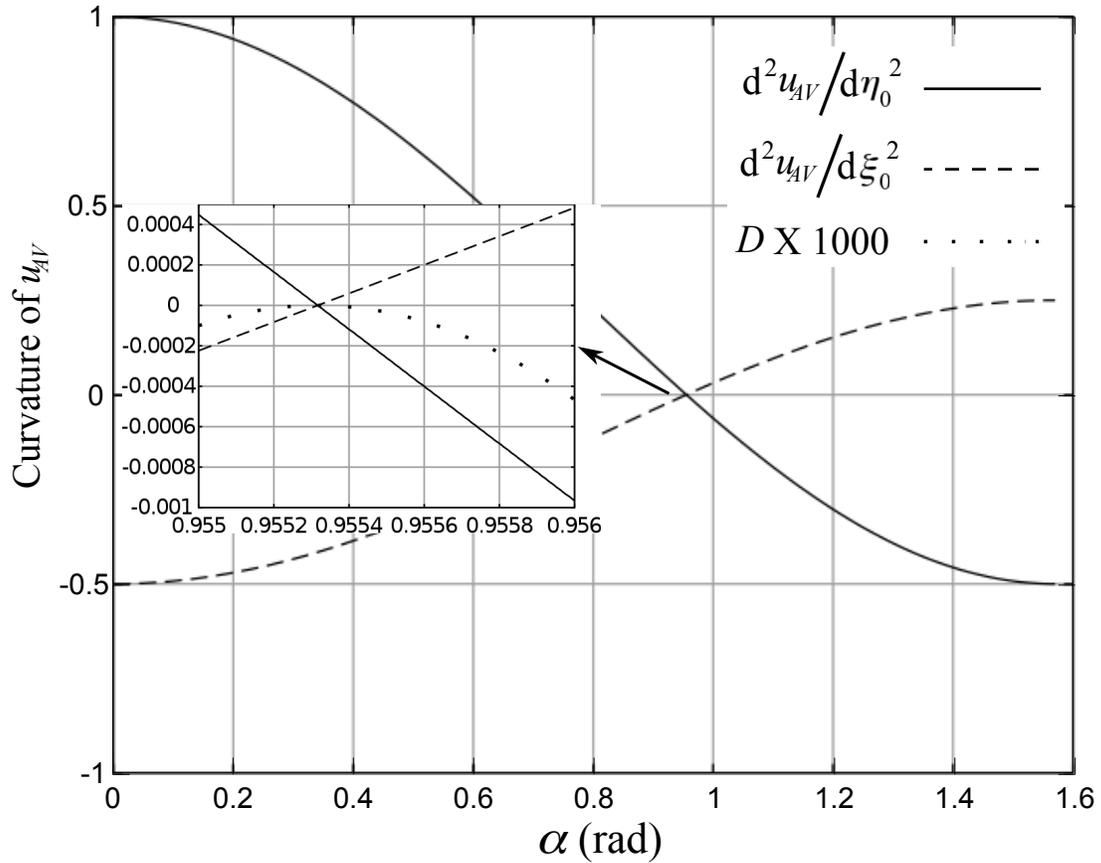

**Figure 4: Numerically Calculated Curvature of the averaged, normalized potential at the coordinate origin as a function of the precessional angle. Inset is an expanded view near where the curves cross and includes the second derivative test discriminant multiplied by 1000.**

In Fig. 4 one can see that the curvature of $u_{AV}$ relative to $\xi_0$ and $\eta_0$ cross and become equal to zero at approximately 0.9553 radians. At this point the second derivative test discriminant, $D \equiv (d^2 u_{AV}/d\xi_0^2)(d^2 u_{AV}/d\eta_0^2) - (d^2 u_{AV}/d\xi_0 d\eta_0)(d^2 u_{AV}/d\eta_0 d\xi_0)$, is also zero to the limits of the numerical procedures used. Note that the angle of ~0.9553 radians (~57.4 degrees) is a root of the second order Legendre Polynomial, $P_2(\cos\theta) = 0$, and also the angle between the interior space diagonal of a cube and the nearest edges of the cube. The angle's exact value is arccos $((1/3)^{1/2})$ and appears widely in science and engineering, often referred to as the magic angle (MA).

Fig. 5 is a surface/contour plot of $u_{AV}$ at the precession angle $\alpha =$ MA.

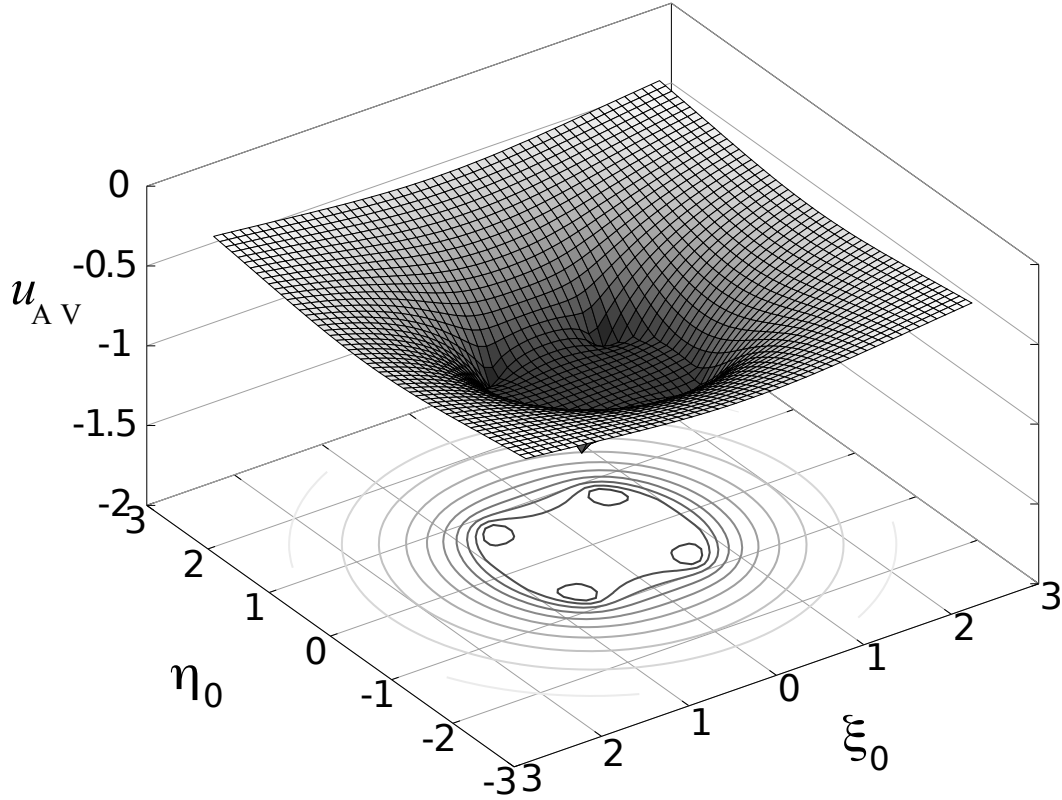

**Figure 5: Surface/Contour plot of the averaged, normalized potential at the magic angle precessional cone half-angle.**

One can see in Fig. 5 that the averaged potential at the MA precession angle is relatively flat inside of the ring radius and that linear motion along the $\xi_0$ and $\eta_0$ axes is stable. However, there are still singularities and the overall shape is still an unstable saddle point. In order to understand how this system can be stable in three dimensions it is necessary to derive and analyze the forces, torques and gyroscopic motion due to gravity.

### III. Force, Torque and Gyroscopic Motion

Note that the origin of the body fixed coordinate system has been set at the center of mass of the ring for mathematical convenience. While the primary interest of this derivation is the motion of the center of mass of the ring, this is fixed at the origin in these coordinates. Thus what is being derived here are the force and torque that will lead to equations of motion for the vector, *X*, from C to the point mass (the motion of the point mass from the point of view of an observer at C in the body fixed frame). The signs on the force and torque are inverted from what one might intuitively expect for an inertial coordinate system centered at the stationary point mass.

Using (9), the formula $\boldsymbol{F} = -\nabla V$, the dimensionless variables $\xi$ and $\eta$, and the normalized force, $\boldsymbol{f} = \boldsymbol{F}/(GM_S M_R/R^2)$, the components of $\boldsymbol{f}$ in the plane of the ring, $f_\xi$ and perpendicular to the plane of the ring, $f_\eta$, can be written in the forms

$$f_\xi = -\frac{\left(1+\xi^2-2\xi+\eta^2\right)\mathrm{K}\left(\frac{4\xi}{1+2\xi+\xi^2+\eta^2}\right) - \left(1-\xi^2+\eta^2\right)\mathrm{E}\left(\frac{4\xi}{1+2\xi+\xi^2+\eta^2}\right)}{\pi\xi\left(1+\xi^2-2\xi+\eta^2\right)\sqrt{1+2\xi+\xi^2+\eta^2}}, \tag{15}$$

and

$$f_\eta = -\frac{2\eta\mathrm{E}\left(\frac{4\xi}{1+2\xi+\xi^2+\eta^2}\right)}{\pi\left(1+\xi^2-2\xi+\eta^2\right)\sqrt{1+2\xi+\xi^2+\eta^2}}, \tag{16}$$

where E is the Complete Elliptic Integral of the second kind.

By defining the unit vector parallel to the $\xi$ axis in body fixed coordinates

$$\boldsymbol{e}_\xi = \frac{(\boldsymbol{X}\cdot\boldsymbol{e}_1)\boldsymbol{e}_1 + (\boldsymbol{X}\cdot\boldsymbol{e}_2)\boldsymbol{e}_2}{\sqrt{(\boldsymbol{X}\cdot\boldsymbol{e}_1)^2 + (\boldsymbol{X}\cdot\boldsymbol{e}_2)^2}}, \tag{17}$$

it is possible to write the dimensionless gravitational force vector in body fixed coordinates,

$$\boldsymbol{f} = f_\xi \boldsymbol{e}_\xi + f_\eta \boldsymbol{e}_3, \tag{18}$$

the dimensionless vector, $\boldsymbol{x} \equiv \boldsymbol{X}/R$, from C to the point mass in body fixed coordinates,

$$\boldsymbol{x} = \xi\boldsymbol{e}_\xi + \eta\boldsymbol{e}_3 \tag{19}$$

and the gravitational torque,

$$\boldsymbol{n} = \boldsymbol{x} \times \boldsymbol{f}, \tag{20}$$

where $\boldsymbol{n}$ is a normalized torque, $\boldsymbol{n} = \boldsymbol{N}/(GM_S M_R/R)$. Since $\boldsymbol{x}$ and $\boldsymbol{f}$ both lie in the plane defined by the $\boldsymbol{e}_3$ and $\boldsymbol{e}_\xi$ axes, the torque, $\boldsymbol{n}$, will always be perpendicular to the $\boldsymbol{e}_3$ axis. The component of the angular momentum parallel to the $\boldsymbol{e}_3$ axis, $I_3\omega_3$, will be a constant of the motion (in the body fixed coordinate system, $\boldsymbol{I}$ is a constant therefore $\omega_3$ is a constant). Note that $\boldsymbol{n}$ can be written as $(\eta f_\xi - \xi f_\eta)\boldsymbol{e}_n$ where,

$$\boldsymbol{e}_n = \boldsymbol{e}_3 \times \boldsymbol{e}_\xi. \tag{21}$$

Using the definition $X = id_0 + kh_0$ and the $\alpha$-$\beta$-3 coordinate system defined in (11), (17) and (21) can be written in terms of the $\alpha$-$\beta$-3 coordinate system:

$$\begin{aligned}
\boldsymbol{e}_\xi &= \frac{(d_0 \cos\alpha \cos\beta - h_0 \sin\alpha)\boldsymbol{e}_\alpha - (d_0 \sin\beta)\boldsymbol{e}_\beta}{\sqrt{(d_0 \cos\alpha \cos\beta - h_0 \sin\alpha)^2 + (d_0 \sin\beta)^2}} \\
\boldsymbol{e}_n &= \frac{(d_0 \sin\beta)\boldsymbol{e}_\alpha + (d_0 \cos\alpha \cos\beta - h_0 \sin\alpha)\boldsymbol{e}_\beta}{\sqrt{(d_0 \cos\alpha \cos\beta - h_0 \sin\alpha)^2 + (d_0 \sin\beta)^2}}
\end{aligned} \tag{22}$$

Define the normalized time, normalized angular velocity, normalized angular momentum and normalized inertia tensor as

$$\begin{aligned}
\tau &= t / \left(\frac{R^3}{GM_S}\right)^{1/2} \\
\boldsymbol{\omega} &= \boldsymbol{\Omega} / \left(\frac{GM_S}{R^3}\right)^{1/2} \\
\boldsymbol{l} &= \boldsymbol{L} / \left(M_R R^2 \left(GM_S/R^3\right)^{1/2}\right) \\
\boldsymbol{J} &= \boldsymbol{I} / M_R R^2
\end{aligned} \tag{23}$$

respectively.

In order to examine the gyroscopic effects and eventually write the equations of motion, the standard relationship between the time derivative of a vector in a rotating coordinate system and the time derivative in a fixed coordinate system [10],

$$\dot{\boldsymbol{A}}_i = \dot{\boldsymbol{A}} + \boldsymbol{\omega} \times \boldsymbol{A}, \tag{24}$$

where the *i* subscript refers to the time derivative in the inertial frame, will be utilized. Also, the vector form of Euler's equations for a rigid body [10],

$$\boldsymbol{J} \cdot \dot{\boldsymbol{\omega}} + \boldsymbol{\omega} \times (\boldsymbol{J} \cdot \boldsymbol{\omega}) = \boldsymbol{n}, \tag{25}$$

will be utilized.

Note that one can recast the approximation that the precessional period is short compared to the orbital motion time scale as $\dot{\boldsymbol{x}}_i \approx 0$ when one performs an average over a precessional period. Using this and (24) one can write the precession averaged velocity and acceleration of $\boldsymbol{x}$ in the rotating frame as

$$\dot{\boldsymbol{x}} = -\boldsymbol{\omega} \times \boldsymbol{x}$$
$$\ddot{\boldsymbol{x}} = \ddot{\boldsymbol{x}}_i - \underbrace{2\boldsymbol{\omega} \times \dot{\boldsymbol{x}}}_{\text{Coriolis}} - \underbrace{\boldsymbol{\omega} \times (\boldsymbol{\omega} \times \boldsymbol{x})}_{\text{Centrifugal}} - \underbrace{\dot{\boldsymbol{\omega}} \times \boldsymbol{x}}_{\text{Euler}}. \tag{26}$$

Combining (25) and (26) one obtains

$$\ddot{\boldsymbol{x}} = \ddot{\boldsymbol{x}}_i + \underbrace{\boldsymbol{\omega} \times (\boldsymbol{\omega} \times \boldsymbol{x})}_{\substack{\text{acceleration} \\ \text{toward} \\ \text{rotational} \\ \text{axis}}} - \underbrace{\left(\boldsymbol{J}^{-1} \cdot \boldsymbol{n}\right) \times \boldsymbol{x}}_{\substack{\text{torque} \\ \text{induced} \\ \text{precession}}} + \underbrace{\left(\boldsymbol{J}^{-1} \cdot (\boldsymbol{\omega} \times (\boldsymbol{J} \cdot \boldsymbol{\omega}))\right) \times \boldsymbol{x}}_{\substack{\text{free} \\ \text{precession}}}. \tag{27}$$

Thus an observer in the rotating frame will see $x$ moving but not changing magnitude over the time scale of one precessional period ($\dot{\boldsymbol{x}}_i \approx 0$ assumption) and infer a fictitious acceleration towards the axis of rotation and a fictitious acceleration due to the free precession. The inertial acceleration (due to gravity) and torque-induced precession would require a large number of free precessional periods to observe under the $\dot{\boldsymbol{x}}_i \approx 0$ assumption.

Using (11), (13) and (22), one can compute the free precession averaged acceleration in the rotating frame due to the torque-induced precession from (27). Figure 6 is an overlay of average torque-induced precessional acceleration vectors on a surface plot of the averaged, normalized potential, $u_{\text{AV}}$, for $\alpha = \text{MA}$.

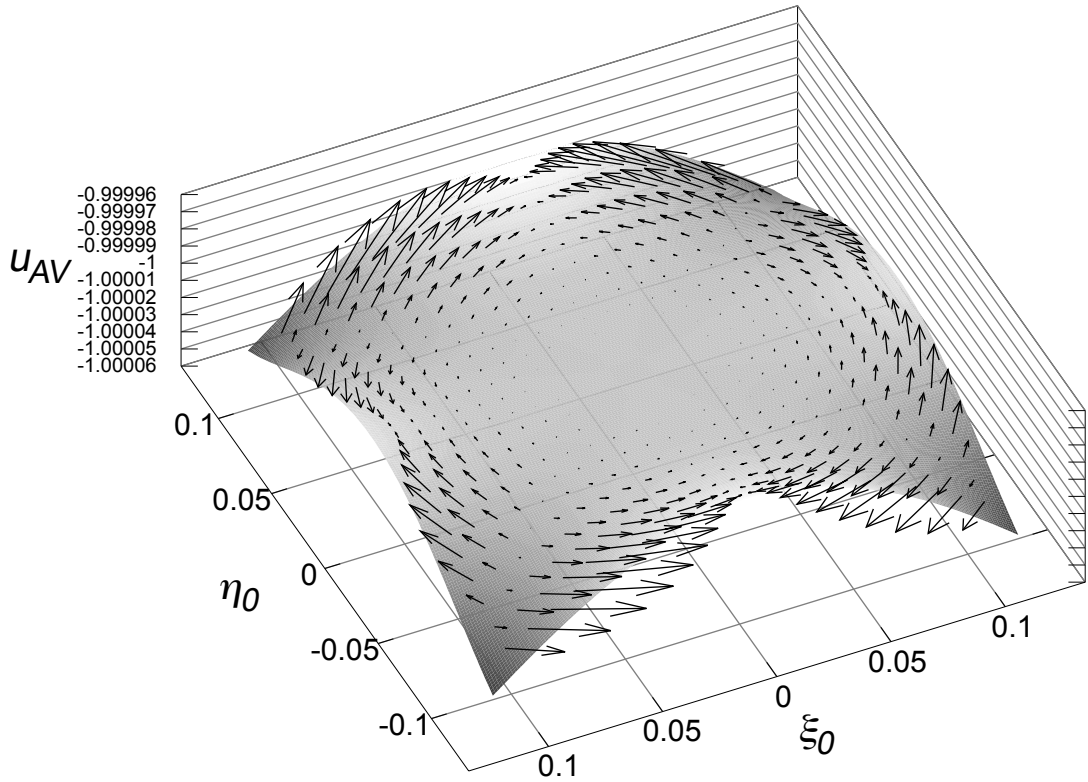

**Figure 6: Quiver plot of torque-induced precessional acceleration, averaged over one free precessional period, overlaid on a surface plot of the center portion of the averaged gravitational potential. Arrows representing precessional acceleration are scaled relative to the axes to increase visibility.**

Figure 6 shows how gyroscopic forces in the ring - point mass system act to make orbits more stable. The torque-induced precession term in the Euler acceleration acts to stabilize the orbits. Note how the arrows in Fig. 6 point away from the unstable parts of the averaged potential, in the valleys along the diagonals, and show a general flow towards the stable ridges along the $\eta_0$ and $\xi_0$ axes. Keep in mind that the arrows represent a fictitious acceleration seen by an observer in the rotating frame of the ring. What they show is that gyroscopic action will tend to reorient the spinning-precessing ring towards a more stable configuration with respect to the point mass it is orbiting. This torque-induced gyroscopic reorientation, along with the potential averaging of the free precession, is what can yield stable orbits.

In an inertial frame, the gyroscopic torque induced precession will become larger at lower spin rates, but the assumption that the precessional period is short compared to the orbital motion time scale, $\dot{x}_i \approx 0$, becomes less accurate as the spin rate decreases. Using this analysis, a possible region where stable orbits may exist is

$$\alpha \approx \arccos\left(\sqrt{\frac{1}{3}}\right)$$
$$\omega_3 \approx 1$$
(28)

The spin was chosen approximately equal to 1 in (25) because stability in three dimensions requires the torque induced precession to be strong enough that it can overcome the instabilities in the averaged potential shown in Fig. 5, but still have a high enough rate of free precession such that the free precession averaging of the gravitational potential is still effective. Since the normalized angular velocity of 1 is equal to the angular velocity of a point mass in a circular orbit of radius $R$, $\omega_3 = 1$ sets the free precessional period to be on the order of the orbital motion.

In order to determine if and under what parameters there are stable orbits, the analysis shall now proceed to a full dynamical simulation of the ring - point mass system.

## IV. Dynamical Simulation

It is convenient to perform dynamical simulations on rigid bodies in a coordinate system fixed to the bodies' principal axes of rotation (the *1-2-3* coordinate system defined in Fig. 1).

The relative orientation of the body fixed, *1-2-3*, coordinate system and a spatially fixed coordinate system will be represented by a unit quaternion of rotation [11], $q_{ROT}$. A quaternion is an ordered quadruplet that can be considered to have a scalar and vector component,

$$q \equiv (q_0, q_1, q_2, q_3) = (q_0, \boldsymbol{q}).$$
(29)

The sum and product of two quaternions are defined as

$$q + p \equiv (q_0 + p_0, \boldsymbol{q} + \boldsymbol{p})$$
$$q \circ p \equiv (q_0 p_0 - \boldsymbol{q} \cdot \boldsymbol{p}, q_0 \boldsymbol{p} + p_0 \boldsymbol{q} + \boldsymbol{q} \times \boldsymbol{p})$$
(30)

respectively and the conjugate and norm of a quaternion are defined as

$$\bar{q} \equiv (q_0, -\boldsymbol{q})$$
$$|q| \equiv \sqrt{q_0^2 + \boldsymbol{q} \cdot \boldsymbol{q}}$$
(31)

respectively. If a point has space fixed coordinates $\boldsymbol{x}$ and body fixed coordinates $\boldsymbol{x}'$, then the quaternions $x$ and $x'$ can be associated with them:

$$\begin{aligned} x &= (0, \boldsymbol{x}) \\ x' &= (0, \boldsymbol{x}') \end{aligned} \tag{32}$$

If the body fixed coordinates are rotated by an angle $\mu$ around an axis of rotation parallel to the unit vector $\boldsymbol{e}_\mu$, $q_{ROT}$, defined as

$$q_{ROT} = \left(\cos\left(\frac{\mu}{2}\right), \sin\left(\frac{\mu}{2}\right) \boldsymbol{e}_\mu\right), \tag{33}$$

relates $\boldsymbol{x}$ and $\boldsymbol{x}'$ by

$$\begin{aligned} x &= q_{ROT} \circ x' \circ \bar{q}_{ROT} \\ x' &= \bar{q}_{ROT} \circ x \circ q_{ROT} \end{aligned} \tag{34}$$

In the body fixed, *1-2-3*, coordinates, the normalized rotational inertia tensor, $\boldsymbol{J}$, is diagonal with $J_{11} = J_{22} = 1/2$ and $J_{33} = 1$.

The equations of motion in the body fixed coordinates, *1-2-3*, can be written as

$$\begin{aligned} \dot{\boldsymbol{v}} &= \boldsymbol{f} - \boldsymbol{\omega} \times \boldsymbol{v} \\ \dot{\boldsymbol{x}} &= \boldsymbol{v} - \boldsymbol{\omega} \times \boldsymbol{x} \\ \dot{\boldsymbol{\omega}} &= \boldsymbol{J}^{-1} \cdot (\boldsymbol{n} - \boldsymbol{\omega} \times (\boldsymbol{J} \cdot \boldsymbol{\omega})), \\ \dot{q}_{ROT} &= \frac{1}{2} q_{ROT} \circ \omega \end{aligned} \tag{35}$$

where the quaternion $\omega = (0, \boldsymbol{\omega})$, $\boldsymbol{v}$ is the time derivative of $\boldsymbol{x}$ in an inertial frame, the dots indicate a time derivative in the rotating, body-fixed frame and $\boldsymbol{f}$ and $\boldsymbol{n}$ are the normalized force and torque from (18) and (20), respectively.

Equations (35) are a coupled set of first order, ordinary differential equations that are straightforward to solve numerically. In order to avoid a build up of numerical errors in the value of $q_{ROT}$, the quaternion is renormalized after each numerical integration time step using the formula

$$q = q/|q|. \tag{36}$$

Figure 7 provides the results of numerical simulations of equations (35) and (36) across a four-dimensional parameter space,

$$\begin{aligned}
\alpha &= \{0 \text{ to } 1.5 \text{ in steps of } 0.01\} \text{ (rad)} \\
\omega_3 &= \{0 \text{ to } 10 \text{ in steps of } 0.1\} \text{ (rad}/\tau) \\
\|\boldsymbol{v}_0\| &= \{10^{-4}, 10^{-3}, 10^{-2}, 10^{-1}\} \text{ (}R/\tau) \\
\phi_0 &\equiv \frac{\pi}{2} - \arccos\left(\frac{\boldsymbol{v}_0 \cdot \boldsymbol{L}_0}{\|\boldsymbol{v}_0\| \, \|\boldsymbol{L}_0\|}\right) = \left\{0 \text{ to } \frac{\pi}{2} \text{ in steps of } \frac{\pi}{20}\right\} \text{ (rad)}
\end{aligned} \qquad (37)$$

where $\boldsymbol{v}_0$ is the initial velocity of the ring relative to the point mass, set up as a perturbation on the orbit, and $\boldsymbol{L}_0$ is the initial angular momentum. Figure 7 presents the maximum deviation of the center of the ring from the point mass position, in units of $R$, for the union of all $\phi_0$ at each $\alpha$ and $\omega_3$. A separate plot is provided for each $\|\boldsymbol{v}_0\|$ as a function of $\alpha$ and $\omega_3$. The initial position of the ring is centered on the point mass. Each point in the 4-parameter-space represents a simulation time of 100 pseudo-years (PYr), where a PYr = $2\pi \, (R^3 / GM_S)^{1/2}$. The simulation ends if the deviation approaches the value one ($R = 1$ in the normalized variables utilized). Thus, if the system is unstable for any value of $\phi_0$ at a particular $\alpha$, $\omega_3$ and $\|\boldsymbol{v}_0\|$, then that pixel on the appropriate plot in Fig. 7 will be equal to one. Only if the system is stable for all values of $\phi_0$ will the pixel have a value less than one.

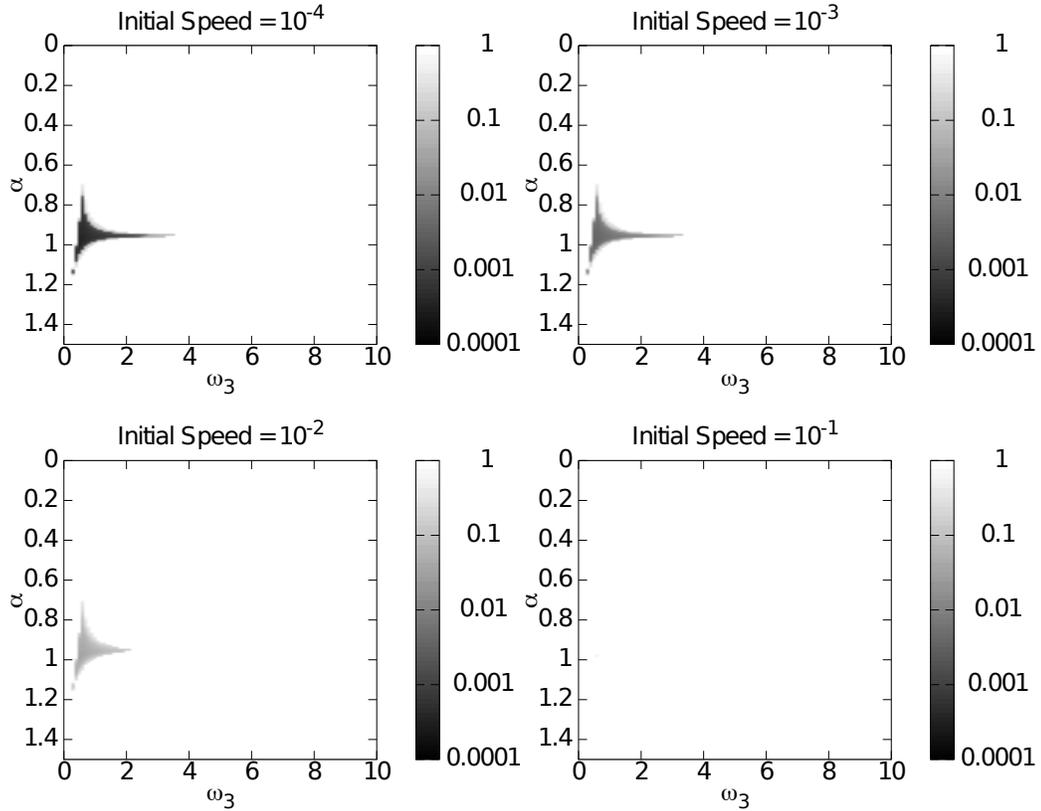

**Fig. 7: Maximum deviation, in units of *R*, of the center of a precessing ring from a point mass as a function of the precession cone half angle, *α*, in radians, the spin, $\omega_3$, in radians/*τ*, and the initial speed in units of *R/τ*, over a range of initial perturbation angles.**

One can see from Fig. 7 that there is a well-defined region of stability in the area of *α* = 0.95 and $\omega_3$ = 1, in line with the expectation from the theoretical analysis. The stability region is approximately triangular with a base near the $\omega_3$ = 0.5 line from approximately *α* = 0.8 to just approximately *α* = 1.1 with a vertex near the *α* = MA, $\omega_3$ = 2 point. The stability holds and the maximum deviation scales with $\|\mathbf{v}_0\|$ up to a value of 0.01. In the lower right of Fig. 7, the stability region disappears at a value of $\|\mathbf{v}_0\|$ = 0.1.

The stability region in Fig. 7 is relatively large in the sense that the stable values of the *α* parameter span a significant portion of its possible values, and the stable values of the $\omega_3$ parameter span a significant range when compared to Keplerian orbital speeds. The stable *α* values span approximately 17 deg, at the base of the triangular stability region, about 10% of its 180 deg range. Comparing the spin, $w_3$, to the orbital speed of a standard Keplerian orbit, the eccentricity, *ε*, of the orbits (considering *R* as the perigee of an Earth orbit) would range from a circular orbit with *ε* = 0 to an unbound, hyperbolic orbit with *ε* = 3. Considering *R* as the apogee of an Earth orbit, the eccentricity ranges from $\varepsilon = 0$ to $\varepsilon = 0.75$. For $R$ (as apogee) equal to the radius of a geosynchronous obit, the perigee of an orbit with $\varepsilon = 0.75$ would be below the surface of the Earth.

Figure 8 presents 3D line plots of precessing ring orbits for an initial speed of the perturbation of 0.01 at various initial velocity angles for a simulation time of 1000 PYr. In Fig. 8, the position of the center of the ring is plotted on a set of spatially fixed axes centered on the point mass with the z-axis set parallel to the initial angular momentum of the ring, $L_0$. One can see from Fig. 8 that precessing ring orbits resemble Lissajous orbits and have an approximate cylindrical symmetry about the $\eta_0$ axis (parallel to the angular momentum). There is no indication of instability in the orbits, and the orbital paths appear very regular over the 1000 PYr time scale. Also, the maximum deviation of the center of the ring from the point mass is approximately the same for each orientation of the initial perturbation velocity.

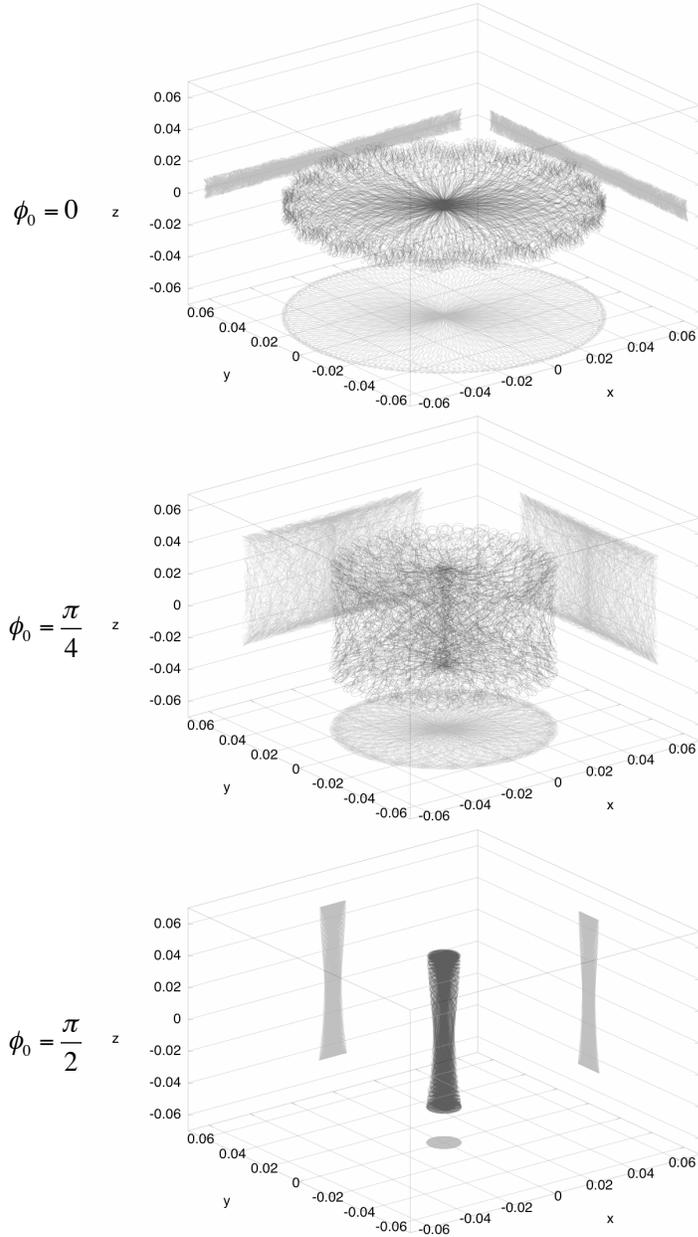

**Figure 8: 3D line plots of precessing ring orbits in space fixed Cartesian coordinates centered on the point mass with the z-axis parallel to $L_0$. Planar projections of the orbit are plotted on each "wall" of the bounding box. All orbits represent 1000 PYr of simulation time with initial conditions of $\alpha = \arccos\left((1/3)^{1/2}\right)$, $\omega_3 = 1$, $\|v_0\| = 0.01$, and top: $\phi_0 = 0$, middle: $\phi_0 = \pi/4$, and bottom: $\phi_0 = \pi/2$.**

The general form of the effective potential that determines the orbits of the precessing ring – point mass system can be outlined using the methods used to generate Fig.'s 7 and 8. Ignoring the internal rotational energy of the ring, the effective kinetic energy of the system is set equal to the linear kinetic energy due to the perturbation velocity. Inspecting the simulation results for the maximum deviation of the ring center from the point mass, the turning points of the motion are approximated and a point on the effective potential energy curve is obtained. Figure 9 shows the results of such an analysis where Fig. 9 (a) represents the full potential at three initial velocity angles and Fig. 9 (b) shows that the effective potential is well fit by a quadratic function at low maximum displacements (displacement $\lesssim$ 10% of $R$). The inset in Fig. 9 (b) shows a 1000 PYr simulation of a precessing ring – point mass system when a constant force along the x-axis is applied. The orbit is stable and compact around the equilibrium point predicted by a Hook's law potential based upon the fitted constant from Fig. 9 (b). Thus the approximation of the effective potential as a Hook's law system is consistent with the simulation. One can see in Fig. 9 (a), that the effective potential is not well behaved at larger displacements (> 20% of $R$) and is not stable against perturbations that cause a displacement larger than approximately 30% of $R$.

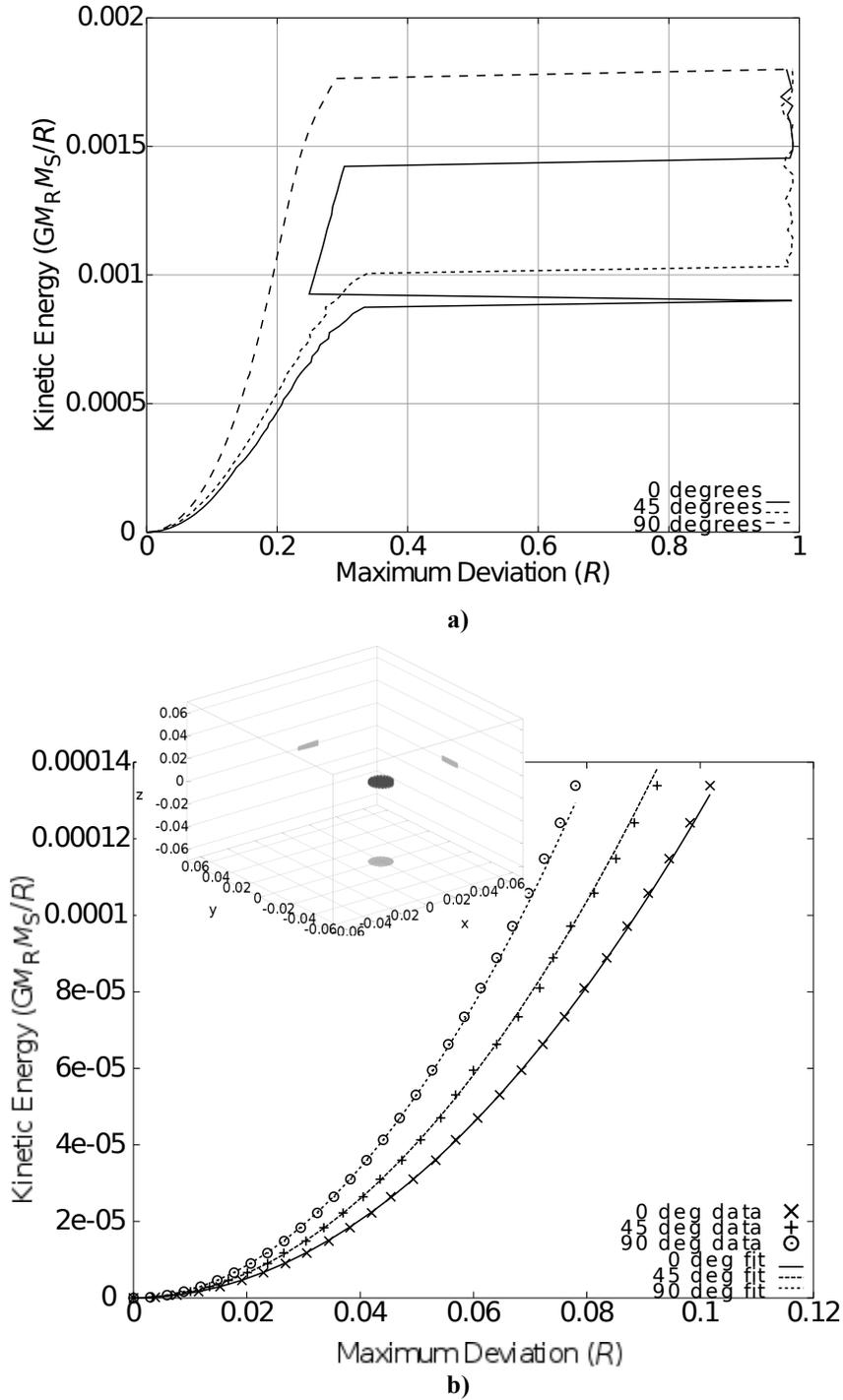

**Fig 9: a)** Effective potential of precessing ring – point mass system as approximated by the maximum deviation of ring center from the point mass. **b)** Close-up of the region below approximately 10% of the ring radius with quadratic (Hook's law) function fit. Inset shows a stable orbit under constant external force along the x-axis.

## V. Notes on Practical Implementation of an Orbital Scale, Rigid Precessing Ring

The analysis presented has assumed a perfectly rigid, massive ring. No material is perfectly rigid, and the long, very thin nature of any likely orbital ring around a celestial body of any significant size would be expected to be extremely flexible for any known material. In this section one approach to increasing the rigidity of a ring is introduced with some relatively simple analysis. Detailed analysis of material properties or engineering details for such a construct are beyond the scope of this paper.

In order to increase the rigidity of a flexible cable in a circular ring, one can increase the tension on the cable by applying a uniform force radially outward. Consider the centrifugal acceleration (acceleration corresponding to the fictitious centrifugal force felt in the body fixed reference frame of the ring) on the ring due to its rotational motion. The rotational motion of the ring can be broken down to a component due to the spin of the ring about the $e_3$ axis, at the angular rate $\omega_3$, and a component in the plane of the ring along what shall be defined as the $e_\omega$ axis at the angular rate $\omega_\omega$. For a given $\alpha$ and $\omega_3$ of the precessing ring one can write

$$\omega_\omega = 2\omega_3 \tan \alpha. \tag{38}$$

The standard analysis of the motion of a freely rotating/precessing, rigid symmetrical body in body fixed coordinates shows that $\omega_\omega$ is a constant, and $e_\omega$ rotates in the plane of the ring at a constant rate [10]. Thus it is possible to consider the instantaneous centrifugal acceleration (due to the fictitious centrifugal force in rotating body fixed coordinates) when $e_\omega$ is parallel to $e_2$ without loss of generality. Using this, along with the conditions of (25), and the formula for centrifugal acceleration at position $r$,

$$\boldsymbol{a}_{CF} = -\boldsymbol{\omega} \times (\boldsymbol{\omega} \times \boldsymbol{r}). \tag{39}$$

The components of $a_{CF}$ in cylindrical coordinates with $z = e_3$ are plotted in Fig. 10.

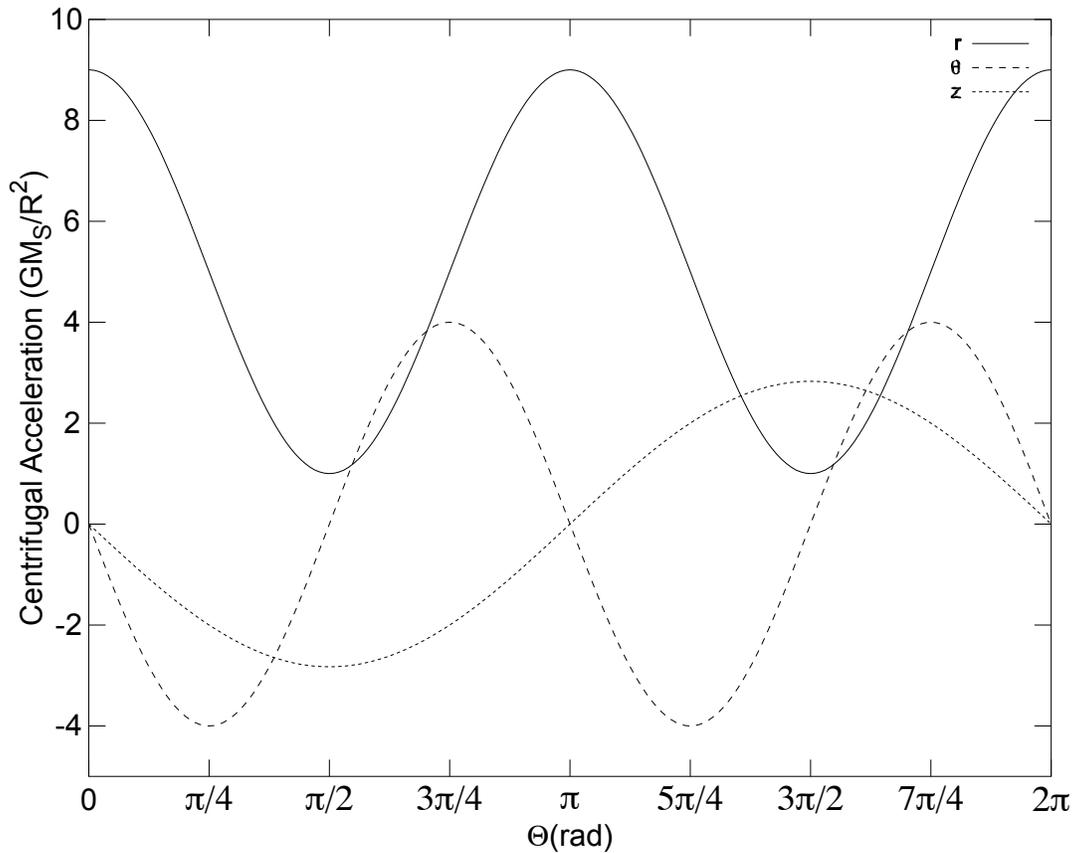

**Figure 10: Components of the centrifugal acceleration of the rigid precessing ring in cylindrical coordinates as a function of the angle around the ring.**

In Fig. 10, the acceleration is not uniform and has components other than radial. Thus, without an additional force to increase the rigidity, a flexible precessing ring may be unstable and might be pulled out of a ring shape by its own rotational motion. A ring made of highly rigid material with a significant cross-sectional area (or other engineering design to increase rigidity) may be stable, but we only consider the limiting case of a very thin, very flexible ring here. In the normalized units used here, the acceleration due to the gravitational attraction of the central point mass has a magnitude of one at the ring when it is centered on the point mass. Since the maximum centrifugal force on the ring is approximately one order of magnitude greater than the gravitational force on the ring, an approximate minimum value of any stabilizing radial force is two orders of magnitude greater than the gravitational force.

Various methods for applying a uniform radial force to a flexible ring can be imagined, both electromagnetically and mechanically. Only one mechanical method will be considered in this paper. Other methods and their applications will be considered in subsequent papers.

### A. Coaxial Symmetrical Top

A rigid body, such as a rigid ring, with two principal axes of inertia that are equal is called a symmetrical top. Now consider two symmetrical tops attached to each other by massless, frictionless, perfectly stiff bearings such that their principle axes of rotation are aligned but both are free to spin about their symmetry axes independently. Call this a coaxial symmetrical top (CST). A simple, idealized example would be a current carrying ring of superconducting cable inside a circular superconducting tube (a superconducting torus). The Meissner Effect will keep the cable centered within the torus magnetically, with very low friction.

The CST consists of two (or possibly more) symmetrical rigid bodies, denoted by subscripts A and B, with the constraints that their centers of mass are co-located, $\omega_{A1} = \omega_{B1}$ and $\omega_{A2} = \omega_{B2}$ in the body fixed *1-2-3* axes used here. In terms of linear motion a CST will behave as a single rigid body. In terms of angular motion, the angular momentum is

$$
\begin{aligned}
L_1 &= (I_{A1} + I_{B1})\omega_1 \equiv I_1\omega_1 \\
L_2 &= (I_{A1} + I_{B1})\omega_2 \equiv I_1\omega_2 \\
L_3 &= I_{A3}\omega_{A3} + I_{B3}\omega_{B3}
\end{aligned}
\qquad (40)
$$

In a system with no torque about the $e_3$ axis, $\omega_{A3}$ and $\omega_{B3}$ are constants as is $L_3 = (I_{A3}\omega_{A3} + I_{B3}\omega_{B3})$. Solving for the motion in body fixed coordinates of either of the coupled rigid bodies will yield the solution for the entire CST. In body fixed coordinates in A, the equations of motion are

$$
\begin{aligned}
I_1\dot{\omega}_1 + \omega_2 L_3 - \omega_{A3} I_1 \omega_2 &= N_1 \\
I_1\dot{\omega}_2 + \omega_{A3} I_1 \omega_1 - \omega_1 L_3 &= N_2 \\
\dot{\omega}_{A3} &= 0 \\
\dot{\omega}_{B3} &= 0
\end{aligned}
\qquad (42)
$$

In the torque free case in body fixed coordinates in A, the angular momentum vector uniformly precesses about the $e_3$ axis at angular rate $\beta_A = L_3/I_1 - \omega_{A3}$. Similarly, in body fixed coordinates in B, $\beta_B = L_3/I_1 - \omega_{B3}$. Comparing this to the standard precession frequency in body fixed coordinates for a simple symmetrical top, $\beta = L_3/I_1 - \omega_3$, it is clear that the free precession behaves identically to a simple symmetrical top where the component of angular momentum about the symmetry axis is replaced by the sum of the angular momenta of the two bodies in the CST case. Similarly, torque induced precession, since it only effects $\omega_1$ and $\omega_2$ which are constrained to be equal, will also only depend upon $I_1$ and $L$ of the CST in an identical manner to how it behaves in a simple symmetrical top, of which a rigid circular ring is an example.

Now consider the simple example where a thin cable is held by stiff magnetic bearings inside a thin torus. Call the cable A and the torus B and compare them to a solid ring of the same total mass, $M_R = M_A + M_B$, and with spin $\omega_3$. Taking $e_\omega$ parallel to $e_2$ as in Fig. 10, and using (40),

$$\begin{aligned}\omega_{A1} &= \omega_{B1} = 0 \\ \omega_{A2} &= \omega_{B2} = 2\omega_3 \tan\alpha \\ \omega_{A3} &= \frac{M_R}{M_A}(\omega_3 - \omega_{B3}) + \omega_{B3}\end{aligned} \quad (43)$$

If $M_A = 0.1 M_R$ and $\omega_{B3} = 0$, then $\omega_{A3} = 10\omega_3$ in order to get the CST to behave identically with the solid ring. However, equation (39) is not the same for the CST and the solid ring. The radial component of the centrifugal acceleration on A due to $\omega_{A3}$ will be 100 times the radial component of the centrifugal acceleration on the solid ring due to $\omega_3$ since (39) depends on the square of the angular speed. Since $M_A = 0.1 M_R$, the total tension force on the CST will be 10 times the tension of the solid ring. Due to the stiff bearing connecting A and B, the resultant tensional stress will be spread across the cross sectional areas of both A and B and will be proportional to the ratio of the masses of A and B if they are both made of the same material. In this case, the tensional stress in the CST will be 10 times the tensional stress in the solid ring if they had equal cross sections. Thus a CST can be designed to provide whatever radial centrifugal force is desired in order to stiffen the ring relatively independently of the forces on the other axes. The non-radial terms in (39) due to the $\omega_\omega$ component of the angular velocity remain but become increasingly negligible as $\omega_{A3}$ becomes larger ($M_A$ becomes smaller). In general for materials with cross sectional area, $A$, density, $\rho$, and Young's modulus, $E$, the stress, $S$, in the materials due to the spins are

$$S_A = \frac{R^2 E_A}{E_A A_A + E_B A_B} \left(A_A \omega_A{}^2 \rho_A + A_B \omega_B{}^2 \rho_B\right)$$
$$S_B = \frac{R^2 E_B}{E_A A_A + E_B A_B} \left(A_A \omega_A{}^2 \rho_A + A_B \omega_B{}^2 \rho_B\right).$$
(44)

If components A and B are made of the same material,

$$S_A = S_B = \frac{R^2 \rho}{A_A + A_B} \left(A_A \omega_A{}^2 + A_B \omega_B{}^2\right).$$
(45)

## B. Practical considerations for a CST

In reality, no bearings are frictionless or perfectly stiff. Even with the advantages of superconductivity with only a heat shield for refrigeration, and the high vacuum of space, the bearings in a CST would be expected to dominate dissipative losses of kinetic energy. In addition, flexing of the structure under the non-radial loads and due to excited vibrational modes of the structure will also cause dissipation. This dissipation will not reduce angular momentum so one would expect a momentum transfer from axes with lower moments to higher moment axes [12]. Primarily this will be a transfer of spin from the lower mass component to the high mass one (from the cable to the torus in the example given). While this transfer will not destabilize the ring orbit directly at first, it will lower the stiffness of the ring and increase the dissipative losses from flexure to transfer angular momentum from the in-plane axes of the ring to its spin. This would move the stability point of the ring up and to the right in Fig. 7. Eventually the ring will either collapse from insufficient radial stiffness or it will exit the stable orbital region in Fig. 7.

It may be possible to design the ring such that its dissipative losses are slow enough to give it a useful lifetime prior to failure, similar to most artificial satellites that require station-keeping thrusters but only carry a limited supply of propellant. However, it is more likely that the ring will be designed with active compensation systems for the dissipative losses. The superconducting cable-inside-torus example is not very practical in terms of charging the magnetic field of the cable or spinning it up. Something more along the lines of a magnetic levitation train design with the ability to apply torque and make small adjustments to the axes between the A and B components of the CST is more likely to be practical. Such a system will require energy input, but that may not be a problem given the size of the ring. As an example, a 100 km circumference ring at 1 A.U. from the Sun, with an effective cross section exposed to the sun of 1 m covered with 10 % efficient solar cells would generate approximately 10 MW of power. The circumference of the Earth is approximately 40,000 km, so 100 km would be a relatively small orbital ring.

## VI.     Applications

Potential applications of a rigid, precessing, massive, orbital ring are very briefly mentioned in this section. These applications and others will be discussed in more detail in subsequent papers.

### A. Gravitational Tractor

From Fig. 9 one can derive that an orbital ring can remain bound to its primary under constant forces up to approximately $10^{-2} \, GM_R M_S / R^2$. If the ring has an electric charge built up on it or a constant current flowing around it, it would catch the solar wind in the manner of a solar electric sail [13] or magnetic sail [14] respectively. The thrust generated from this could be used to change the orbit of the primary the ring is orbiting similar to suggestions in [15] for moving asteroids. The concept in [15] could provide more thrust, but the Hooke's law effective potential of the orbital ring would remove the necessity of having a feedback control system to keep the ring bound to its primary. This may prove most useful when applied to asteroid mining, as the ring could act as a base to process the asteroid while towing it to the desired location for the end product. Since the thrust is applied gravitationally, an asteroid can be towed even while it is being broken up or melted during processing. Such a construct could also be used to move near Earth asteroids to orbits with less probability for hitting the Earth, given enough lead-time.

### B. Artificial Magnetosphere

An orbital CST could generate a magnetic field. This magnetic field could provide an artificial magnetosphere protecting the primary, the ring, and objects in orbit from particle radiation from the solar wind. A configuration of the current carrying wire to generate the field that approximates a hollow cylinder of current would produce little or no magnetic field inside the cylinder. This would be useful in order to avoid complications for inhabitants or electronics inside the ring. This may be beneficial for situations where long-term exposure to the solar wind would be unavoidable, such as the asteroid mining application from the previous section. In a more speculative application, a much larger magnetized ring could protect an entire planet, such as Mars, which has no strong natural magnetic field.

### C. Polyakov-Ring

As a speculative application, similar to a space elevator [16], a CST precessing orbital ring can be geostationary. The angular momentum vector, and therefore precession axis, of the ring can be aligned with the rotational axis of the Earth (or other primary), and a set of parameters chosen such that the precessional angular speed of the ring is equal to the Earth's rotational angular speed. Utilizing a CST with a spin of $2\omega_3$ on the outer segment, the entire surface of the ring would be stationary with respect to an observer on Earth. The possible orbits for a ring at precession angle MA are obtained by using the stability region of Fig. 7 along with the formula for the precession rate, $\omega_p = L_3/I_1 \cos(\text{MA})$. Geostationary orbits exist from approximately $1.4R_\text{Clarke}$ to $3.6R_\text{Clarke}$, where $R_\text{Clarke}$ is the radius of a standard geostationary orbit, approximately 42,000 km for the Earth. Thus the orbits are outside of the existing geostationary belt and are inclined close to 54.7 degrees to the equator. Such a ring would be easily visible from equator to pole. It is possible to have multiple geostationary rings at various orbital radii and phase angles relative to the precession of the other rings.

### D. Niven-Ring

Even more speculative is the construction of an inhabited CST ring as an analog of a Niven-ring either around a star or planetary body. Consider a two component CST where the A component has a spin that provides a 1g radial centrifugal acceleration, and the B component has zero spin. Using approximate parameters from [2], $M_S = 10^{30} kg$, $M_R = 10^{27} kg$, $R = 1$ A.U. and $\omega_3 = (GM_S/R^3)^{1/2}$, equation (43) yields approximately $M_A = 0.0174 M_R$ or $1.74 \times 10^{25}$ kg. Using (39) to perform a calculation similar to Fig. 10, the maximum deviation of the radial centrifugal acceleration of the A ring from 1g is 2.4%, the maximum ratio of the z axis centrifugal acceleration to the radial axis is 4.9%, and the maximum theta axis acceleration ratio to the radial axis is 0.12%. In terms of habitability, the periodic ~5% tilt in the z direction may cause problems with building construction, motion of bodies of water, etc. Reverting to the analogy of the magnetic bearings with a magnetic levitation train, many high speed trains have a mechanism that tilts the train cars as they go around a corner to keep the interiors level in the frame of the occupants. A similar mechanism could be utilized in a Niven-ring CST structure to maintain a stable direction to the artificial gravity direction relative to the occupied area.

## VII. Conclusions

The orbital stability of rotating, precessing, rigid rings was considered in this paper. A universal stability region based on normalized variables was demonstrated qualitatively by theory and quantitatively by simulation. Extension of this both theoretically, with a stability analysis properly accounting for gyroscopic forces, and numerically, with a more detailed simulation which takes into account flexibility in the ring and a many body system of point masses, would be desirable. The coaxial symmetrical top is a novel mechanical system presented here to aid in practical application of precessing orbital rings. Expanded details of possible applications of CST based orbital rings as well as analysis of electromagnetically rigidized rings, along with their interaction with the solar wind, can also be profitably investigated.